\documentclass[pre,twocolumn]{revtex4}
\usepackage[english]{babel}
\usepackage{amsmath}
\usepackage{amssymb}
\usepackage{epsfig}

\begin{document}

\title{Discrete Vortex Filaments on Arrays of Coupled Oscillators\\
in the Nonlinear Resonant Mode}
\author{Victor P. Ruban}
\email{ruban@itp.ac.ru}
\affiliation {Landau Institute for Theoretical Physics RAS,
Chernogolovka, Moscow region, 142432 Russia}

\date{\today}

\begin{abstract}
Numerical simulation has indicated that vortex structures
can exist for a long time in the form of quantized
filaments on arrays of coupled weakly dissipative nonlinear 
oscillators in a finite three-dimensional domain
under a resonant external force applied at the boundary of the domain. 
Ranges of the parameters of the system and an external signal 
favorable for the formation of modulationally stable quasi-uniform 
energy background, which is a decisive factor for the occurrence 
of this phenomenon, have been qualitatively revealed.
\end{abstract}

\maketitle

It is known that quantized vortices are coherent
structures characteristic of nonlinear complex wavefields 
at the matching of the signs of dispersion and
nonlinearity [1-7]. In particular, the Gross-Pitaevskii
equation (defocusing nonlinear Schr\"odinger 
equation with an external potential) describes vortices 
in trapped Bose-Einstein condensates of cold atoms. 
These objects are actively studied (see, e.g., [8-17]).

In the last decades, in view of the development of
technology of metamaterials (in a wide sense), coherent 
structures are studied not only in continua but also
in (quasi-)discrete systems (discrete solitons and
breathers, vortices and vortex solitons on lattices; see
[18-30] and references therein).

It is noteworthy that only a small fraction of these
studies are devoted to vortices as long-range objects
against the modulationally stable background,
whereas most of these studies concern localized structures
characteristic of modulationally unstable discrete systems. 
In particular, traditional vortices were considered in [24, 27, 28] 
using the discrete nonlinear Schr\"odinger equation (DNSE), 
and vortices on lattices of oscillators were simulated numerically in
recent works [29, 30].

The numerical experiments showed that dissipation 
should be very small (the Q factor of oscillators
should be $Q\gtrsim 10^4$ ) for the observation of the dynamics
of interacting vortex structures in an autonomous discrete 
system before the system passes to the linear regime.

A problem arises: How can this requirement of such
a huge Q factor be weakened? A possible solution to it is
to maintain the energy background of the system by
means of a time-monochromatic pumping. For the
external force not to change the dynamic properties of
the lattice, pumping should be applied only to the sites at
the boundary of the system. For vortices to be ``favorably'' 
formed and, then, to continue their existence
inside the array, the initial phases of driving signals
should be made smoothly varying from one boundary
site to another. Such a recipe was successfully used in
[30], where two-dimensional arrays of nonlinear oscillating 
electric circuits joined by capacitive links to the
united circuit, as shown in Fig.1, were simulated.
Weakly dissipative vortex structures in this model were
observed in numerical experiments for many thousands 
of oscillation periods. The energy background was
quasi-uniform in space because oscillators were in the
nonlinear resonance mode (on its upper branch), when
the amplitude of oscillations is determined primarily by
the frequency of the external periodic force and, to a
minor degree, by its amplitude. However, parametric
domains favorable for vortices were not determined.

\begin{figure}
\begin{center}
\epsfig{file=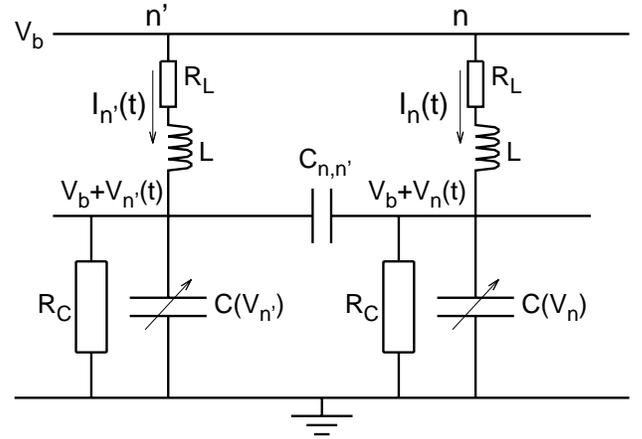, width=90mm}
\end{center}
\caption{Idealized electric circuit corresponding to Eqs. (4)
and (5). Only a fragment of the entire network including
two cells and a link between them is shown.}
\label{scheme} 
\end{figure}

In this work, to make the next natural important
step in studying such systems, we perform extensive
numerical experiments in order to determine parameter 
ranges where the external driving leads to the formation 
of a stable background and to the nucleation of
vortices already in the three-dimensional array. These
are the first results of this kind. As far as I know, the
existence of long-lived vortex filaments on lattices of
oscillators in the nonlinear resonance mode has not
yet been reported. 

To clarify the essence of the problem, it is convenient to first 
consider the DNSE as a simpler model. The weakly dissipative DNSE 
with a monochromatic pump has the form (see, e.g., [31, 32])
\begin{eqnarray}
&&i(\dot A_n+\gamma A_n) = (-\delta+g|A_n|^2)A_n \nonumber\\
&&\qquad\qquad+\frac{1}{2}\sum_{n'}c_{n,n'}(A_n-A_{n'}) +f_n,
\label{A_n_eq}
\end{eqnarray}
where $A_n(t)$  are the unknown complex functions at the
sites $n=(n_1,n_2,n_3)$ of the three-dimensional lattice, $\gamma$
is the low linear damping rate, $\delta$ is the detuning of the
pump frequency from the linear resonance, $g$ is the
nonlinear coefficient, $c_{n,n'}$ is the (real) coupling matrix
(usually between nearest neighbors), and $f_n$ are the
complex amplitudes of the external force.

It is well known that the DNSE is related to various
systems of coupled nonlinear oscillators (see, e.g., [28, 29]) 
because many oscillatory systems in the weakly
nonlinear limit are reduced to the DNSE for complex
envelopes $A_n(t)$ of the canonical complex variables
$a_n=\sqrt{S_n}\exp(i\Theta_n)=A_n(t)\exp(-i[\omega_0+\delta] t)$, 
where $S_n$ and $\Theta_n$ are the action and angle variables for a single
oscillator, respectively, and $\omega_0$ is the frequency of free
oscillations in the limit of small amplitudes. The phase $\Theta$
changing slowly upon the bypass along a closed
contour consisting of the links between neighboring
sites can acquire an increment multiple of $2\pi$, thus
forming a discrete quantized vortex. It is noteworthy
that the variable $S$  does not necessarily vanish at any
site in the core of the vortex. This is a fundamental difference
of discrete vortices from continuous ones.

The capability of the system to support vortices
depends on the relation between the signs of $g$ and $c_{n,n'}$.
These signs should coincide with each other in the
simplest case of identical interactions between the
nearest neighbors on the regular lattice. However, this
condition alone is insufficient for the existence of
long-lived vortex structures in the system. Let a finite
system be composed of oscillators on the cubic lattice
within some three-dimensional domain ${\cal D}$ and all
nonzero parameters $f_n$ have the same amplitude $f$ but
different phases $\varphi_n$. Even ignoring the shape of the
domain, the set of the parameters $\varphi_n$, and amplitude $f$,
there are two significant coefficients $\gamma/\delta$ and $c/\delta$ 
(the coefficient $g/\delta$ in the DNSE can be made unity by
varying the scale of the variable $A$). The picture
appearing with time strongly depends on these parameters. 
It is also important that the parameter $\delta$ and
analogs of the parameters $\gamma$ and $c$ in the initial 
fully nonlinear equations of motion of oscillators are
important individually because there occur parametric 
resonances that are disregarded in Eq.(1) and
destroy the quasi-uniform energy background. In fact,
these are nonlinear wave processes of the $p\rightarrow 2$ type,
where $p$ is the number of decaying waves with zero
quasimomentum. The appearance of such resonances
with an increase in the coupling parameter $c$ and 
corresponding unstable modes are determined by the
specificity of the system. In particular, the properties
of the main parametric resonances for the circuit
shown in Fig.1 will be significantly different depending 
on the type of nonlinear capacitances used. The
answer for the Klein-Gordon lattice
$$
\ddot q_n+2\gamma \dot q_n+q+q^3+\sum_{n'}c_{n,n'}(q_n-q_{n'})
=F_n\cos([1+\delta] t+\varphi_n)
$$
will differ even more strongly because of the significantly 
different dispersion relation of linear perturbations.

It is obviously difficult to scan in detail the ranges
of all parameters in numerical simulations. Consequently, 
it is reasonable at the first stage to fix, e.g., the
shape of the domain, phases $\varphi_n$, and frequency detuning $\delta$. 
Several values should be taken for each of the
remaining parameters $\gamma$, $c$, and $f$. Thus, to obtain a
quite clear picture, it is necessary to perform several
dozen numerical experiments. I performed these
experiments within the electric model shown in Fig.1
rather than within the DNSE. Thus, this study is in the
trend of simulation of basic nonlinear phenomena by
example of electric circuits [33-46]. However, the
qualitative results reported here extend beyond a particular 
model because I also revealed similar vortex
structures on the Klein-Gordon lattice.

We consider an electric network consisting of nonlinear
oscillating circuits with capacitive links between
them shown in Fig.1. The relation between this network
and Eq.(1) was discussed in my recent work [29]. The
state of the system is described by voltages $V_n(t)$ and by
currents $I_n(t)$ through the induction coils $L$. Here, the
capacitances $C(V_n)$ are nonlinear elements. In this
work, two types of the functional dependence of the
capacity on the voltage are used. The first dependence
has the form
\begin{equation}
C(V_n)=C_0(1+V_n^2/V_*^2),
\label{CV_symm}
\end{equation}
where $V_*$ is a parameter. Such a symmetric dependence 
is characteristic of capacitors with dielectric films [47, 48]. 
Another dependence is typical of varactor diodes, 
which (in the presence of a reverse bias voltage $V_b$
and in parallel connection with a normal capacitor)
are approximately described by the formula (see, e.g., [49])
\begin{equation}
C(V_n)=C_0\Big[\mu+(1-\mu)/(1+V_n/V_*)^\nu\Big].
\label{CV_diode}
\end{equation}
Here, the parameter $0<\mu<1$  describes the normal
capacitor connected in parallel and $\nu$ is the fitting
parameter of a varactor diode that is in the range 
$0.3\lesssim \nu\lesssim 6.0$
depending on the fabrication technology. In this work,
we take $\mu=0.5$ and $\nu =2$ .

Formulas (2) and (3), as well as the circuit in Fig.1,
are idealized because a real three-dimensional array
should contain such a large number of sites that it
hardly can be assembled from common electronic elements 
and will be too bulky and expensive. The considered 
idealized circuit more likely describes a certain
miniature three-dimensional structure with similar
electrical properties.

The electrostatic energy stored on the capacitor
(additional to the state with $V_n=0$) is
$$W(V_n)=\int_{0}^{V_n}C(u)u du.$$

The low active resistance of the coil $R_L\ll\sqrt{L/C_0}$
and the high leakage resistance of the capacitor $R_C\gg\sqrt{L/C_0}$
are the dissipative elements of the circuit. 
The dimensionless damping coefficient has the form
$$
\frac{\gamma}{\omega_0}=
\left(R_L\sqrt{C_0/L}+R_C^{-1}\sqrt{L/C_0}\right)/2=(\gamma_L+\gamma_C)/2.
$$
In addition, an alternating voltage ${\cal E}_n(t)$ is applied to
oscillators located at the boundary of the domain. The
corresponding system of equations of motion has the form
\begin{eqnarray}
&&C(V_n)\dot V_n+\sum_{n'} C_{n,n'}(\dot V_n -\dot V_{n'})+V_n/R_C= I_n,
\label{current}\\
&&L \dot I_n \!+\! V_n \!+\! R_L I_n = {\cal E}_n\!(t)
=F_n\!\cos([\omega_0\!+\!\delta]t\!+\! \varphi_n).
\label{voltage}
\end{eqnarray}
In the absence of links and dissipation, the energy
$\varepsilon_n=LI_n^2/2+W(V_n)$ of each oscillator would be conserved.

The calculations were performed with dimensionless 
variables corresponding to the values $L=1$, $C_0=1$, and $V_*=1$. 
In this case, the frequency and
period of small oscillations are $\omega_0=1$ and  $T_0=2\pi$,
respectively. Equations (4) were solved with respect to
$\dot V_n$ using an appropriate iterative procedure and the
fourth-order Runge-Kutta algorithm was used to
obtain the time evolution.

The interior of the ellipsoid $x^2+y^2+2.5z^2<3$ was
taken as the domain ${\cal D}$, and the step of the cubic lattice
$h=0.06$ or $0.04$ determined the total number of
degrees of freedom of the system. The pump signal was
applied to the lattice sites located in the thin shell 
$0<(3-x^2-y^2-2.5z^2)<2h$. 
The pump phases were given by the simple formula 
$\varphi_{n_1,n_2,n_3}=3.0\pi h n_3=3.0\pi z_n$.

\begin{figure}
\begin{center}
\epsfig{file=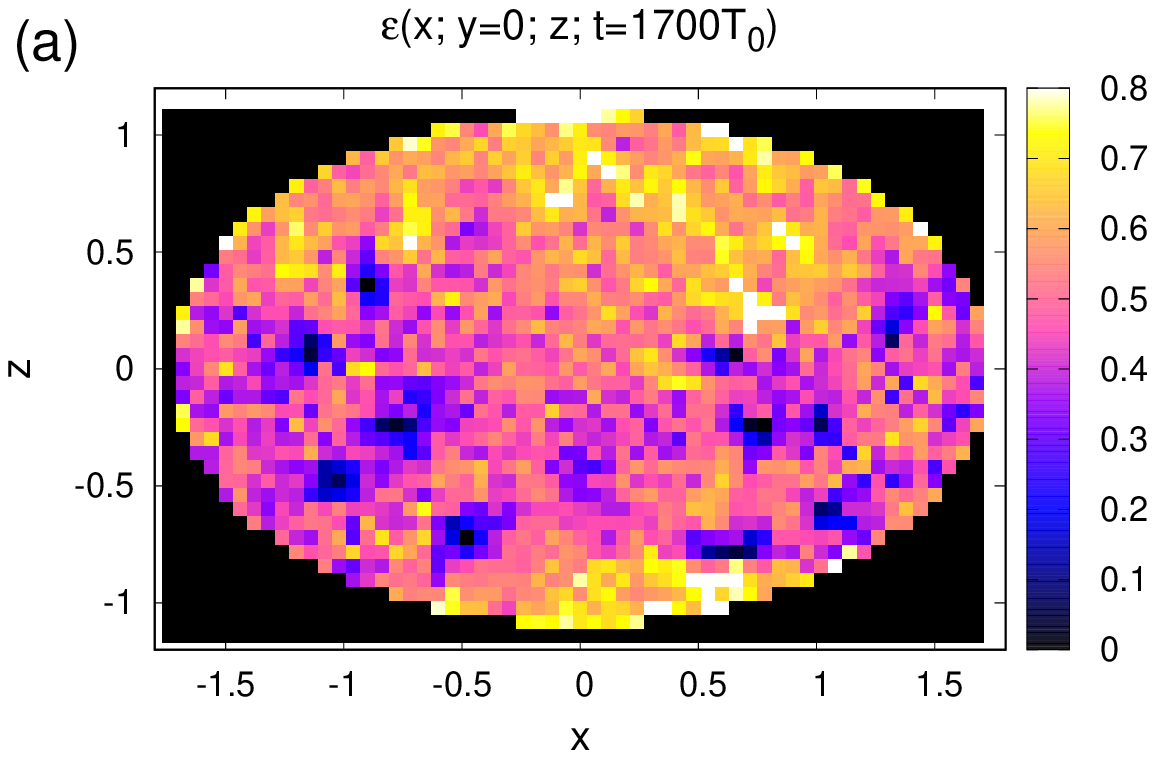, width=75mm}
\epsfig{file=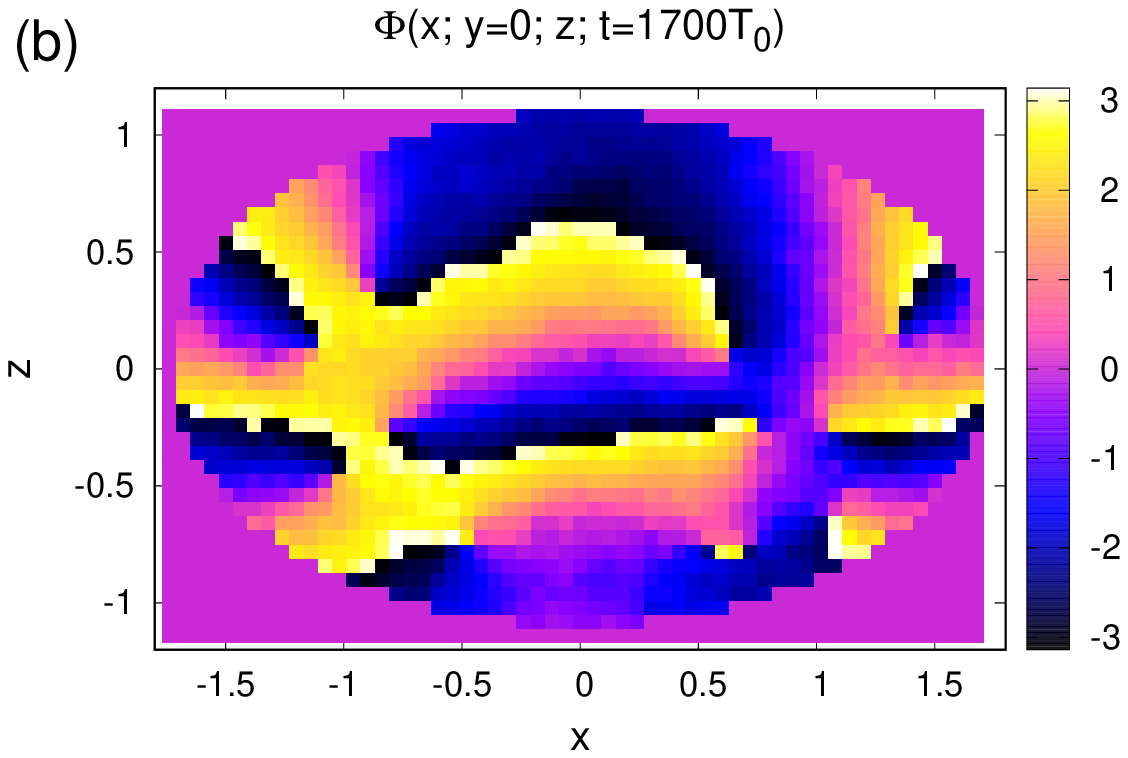, width=75mm}
\epsfig{file=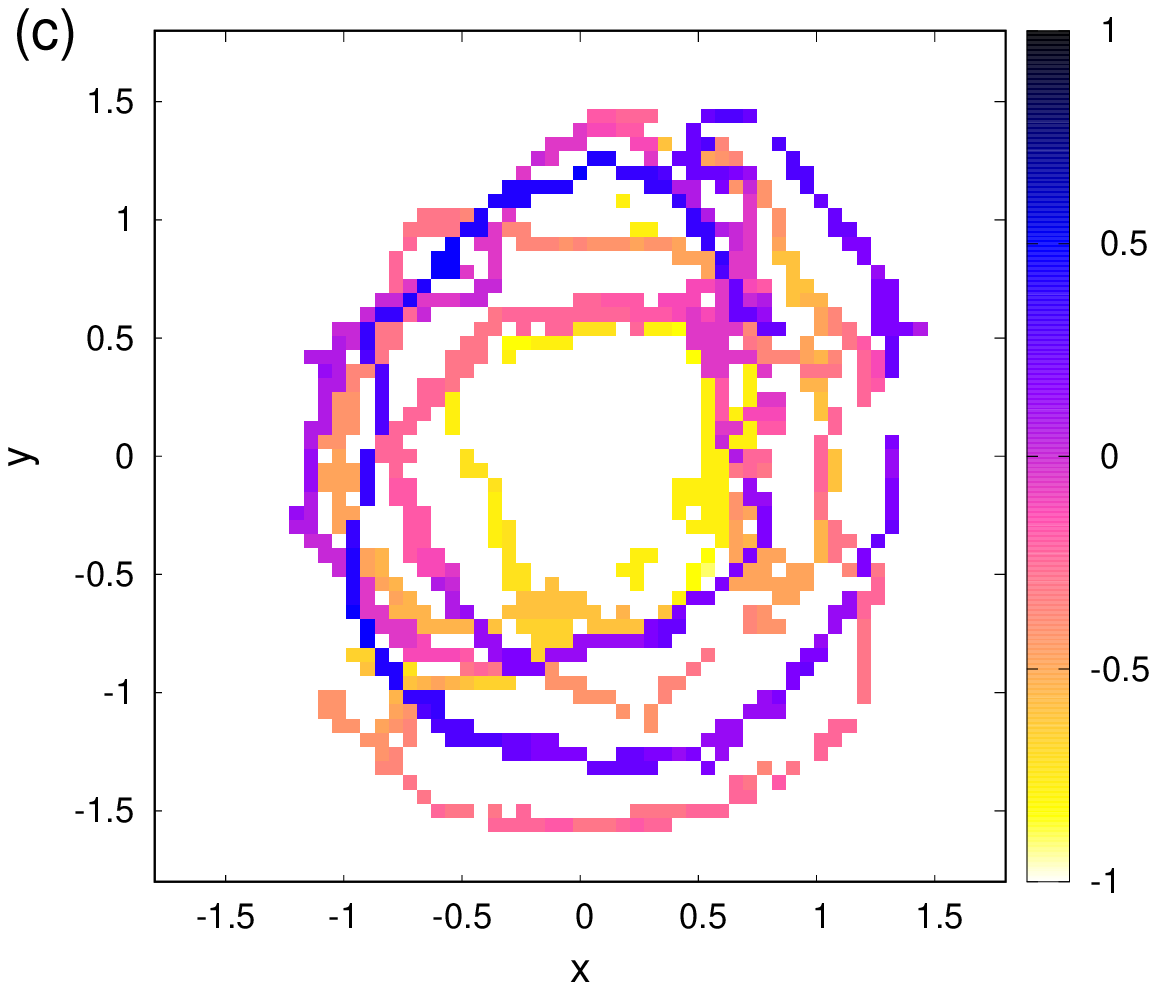, width=75mm}
\end{center}
\caption{
Vortex configuration formed at the time $t=1700 T_0$ and the parameters 
$\gamma_L=\gamma_C=0.001$, $c=0.1$, and $F=0.06$: 
(a) energies of oscillators in the $y=0$ layer, 
(b) their phases, and (c) projection of vortex filaments 
on the $(x,y)$ plane, where the color indicates the
$z$ coordinate of the sites of the lattice where $\varepsilon<0.125$.
Each colored square (``pixel'') corresponds to an individual 
oscillator on the cubic lattice.
}
\label{MU01EQ10C01_06} 
\end{figure}
\begin{figure}
\begin{center}
\epsfig{file=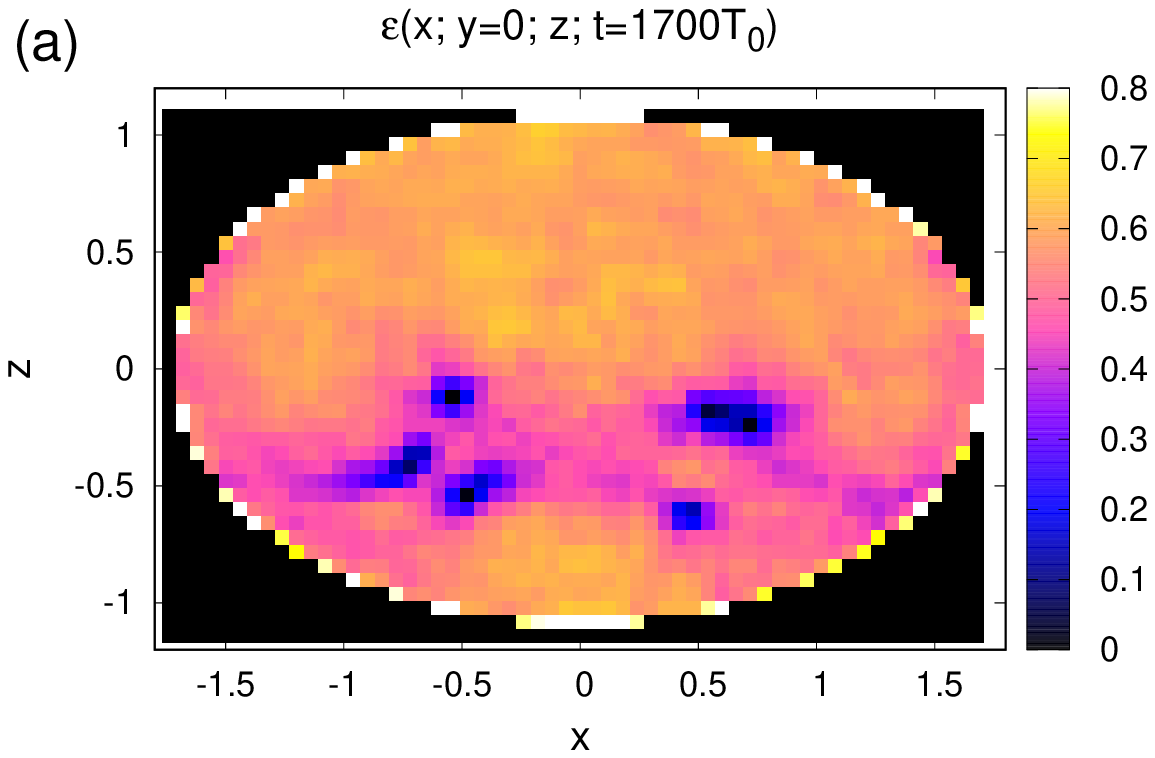, width=75mm}
\epsfig{file=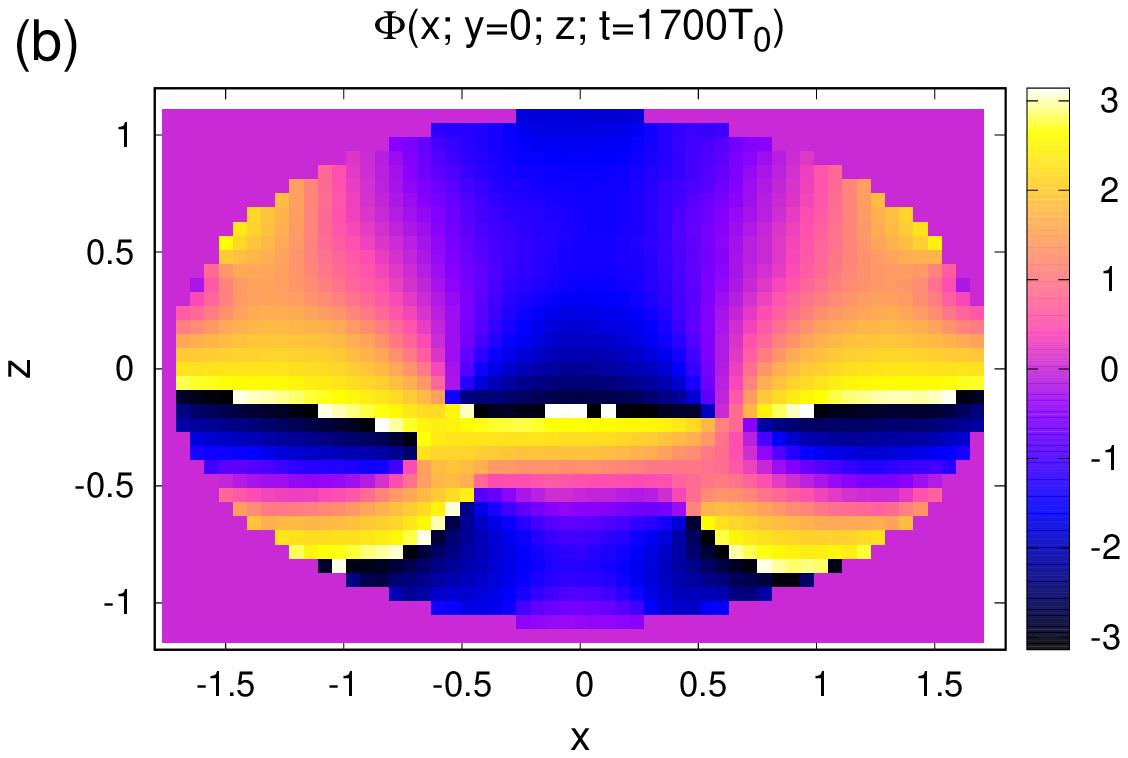, width=75mm}
\epsfig{file=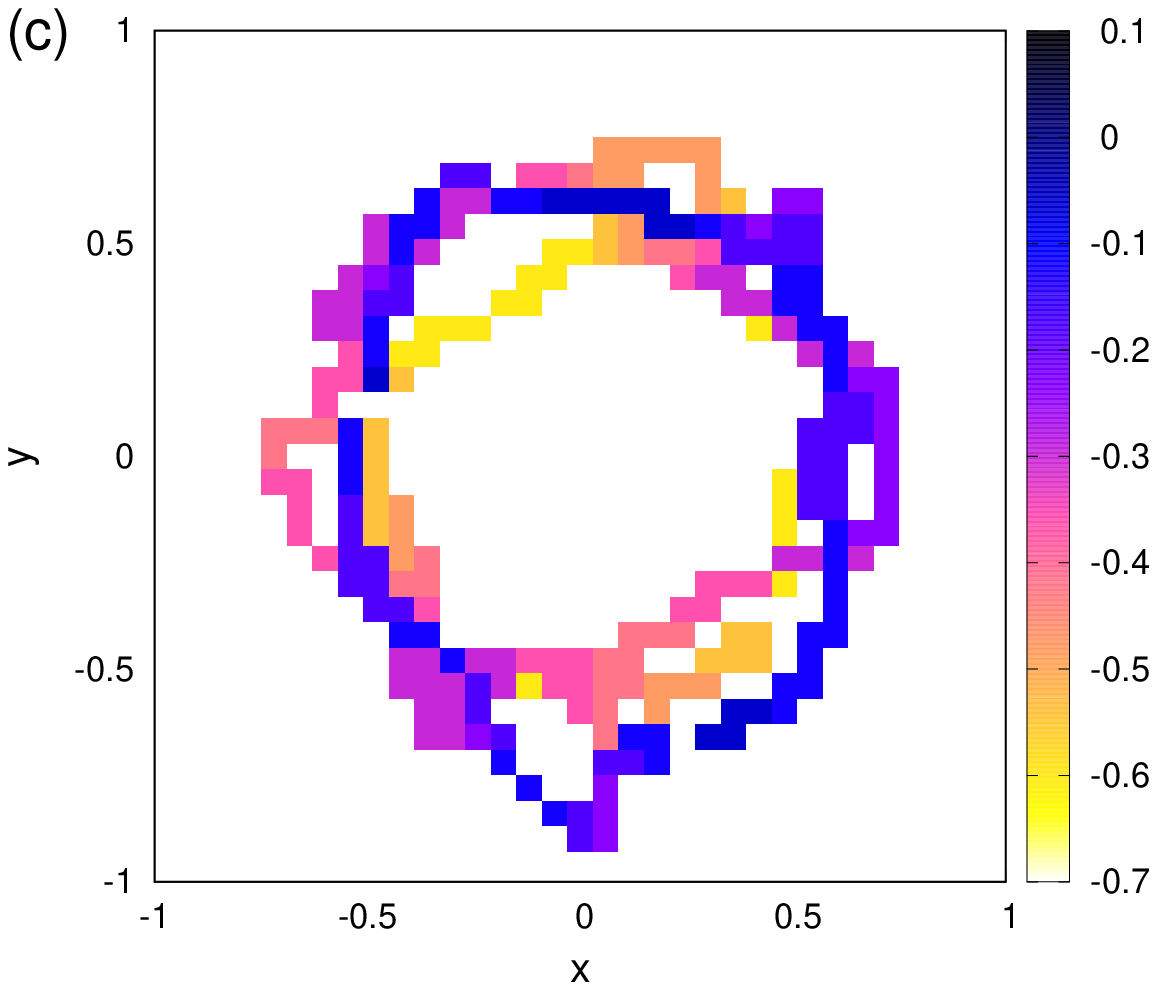, width=75mm}
\end{center}
\caption{
Vortex structure consisting of three rings that is 
formed at the parameters $\gamma_L=\gamma_C=0.001$, $c=0.1$,
and $F=0.12$: (a) energy profile in the $y=0$  layer,
(b) phases of oscillators in this layer, and (c) projection of
vortex filaments on the $(x,y)$ plane.
}
\label{MU01EQ10C01_12} 
\end{figure}
\begin{figure}
\begin{center}
\epsfig{file=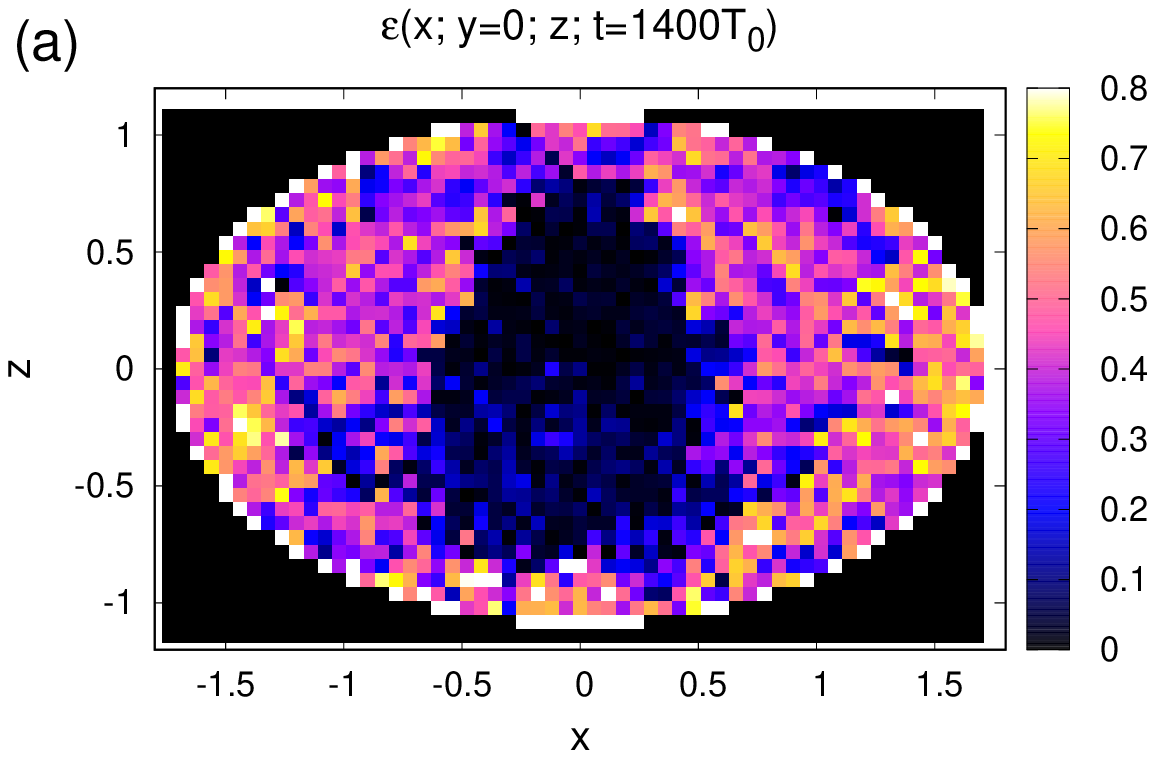, width=75mm}
\epsfig{file=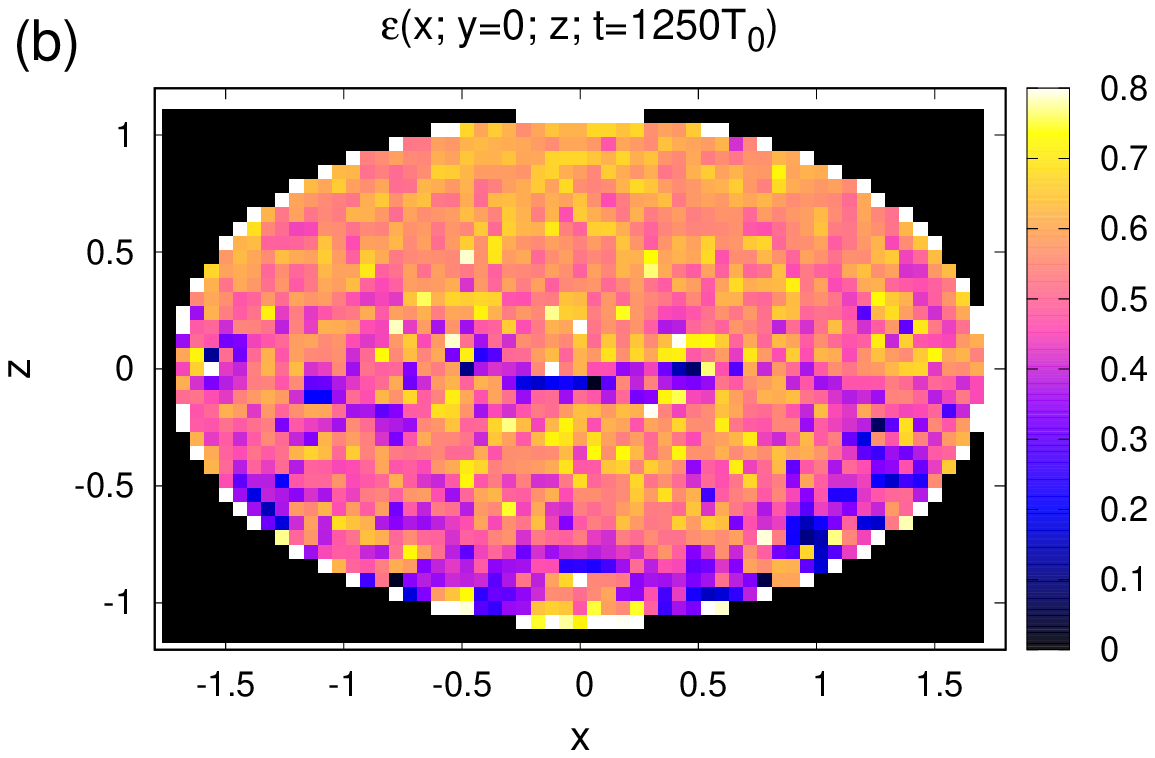, width=75mm}
\epsfig{file=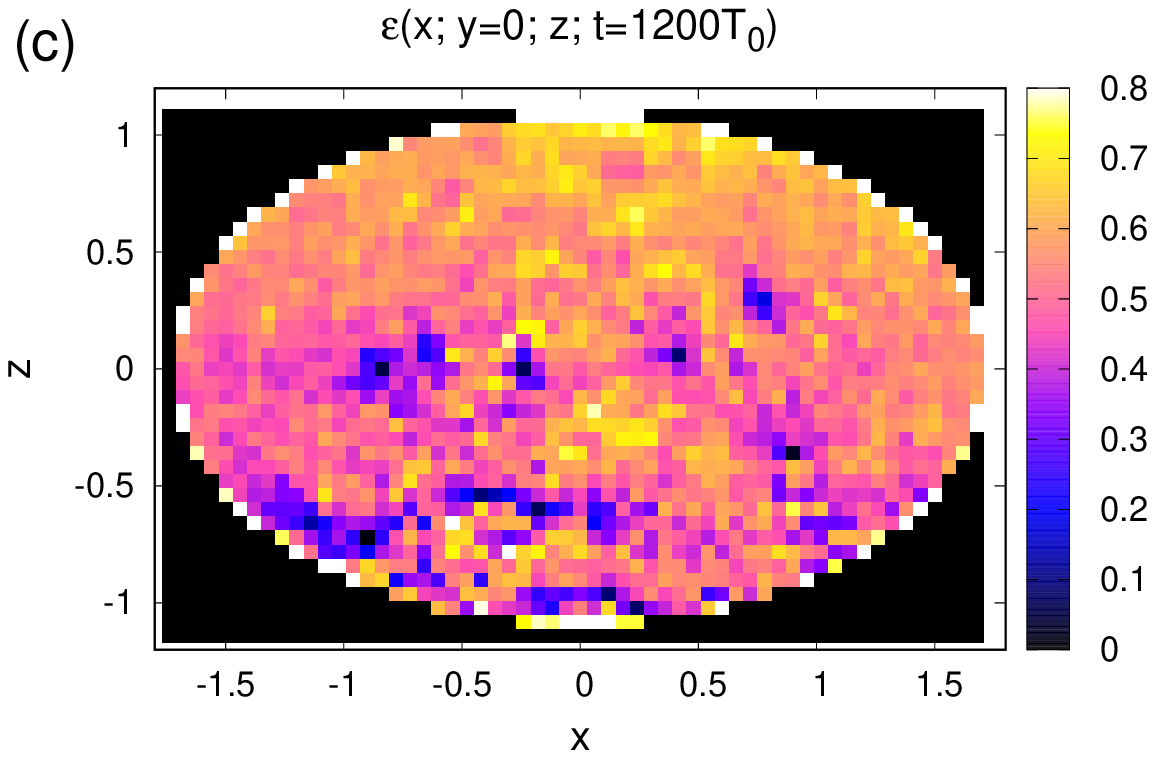, width=75mm}
\epsfig{file=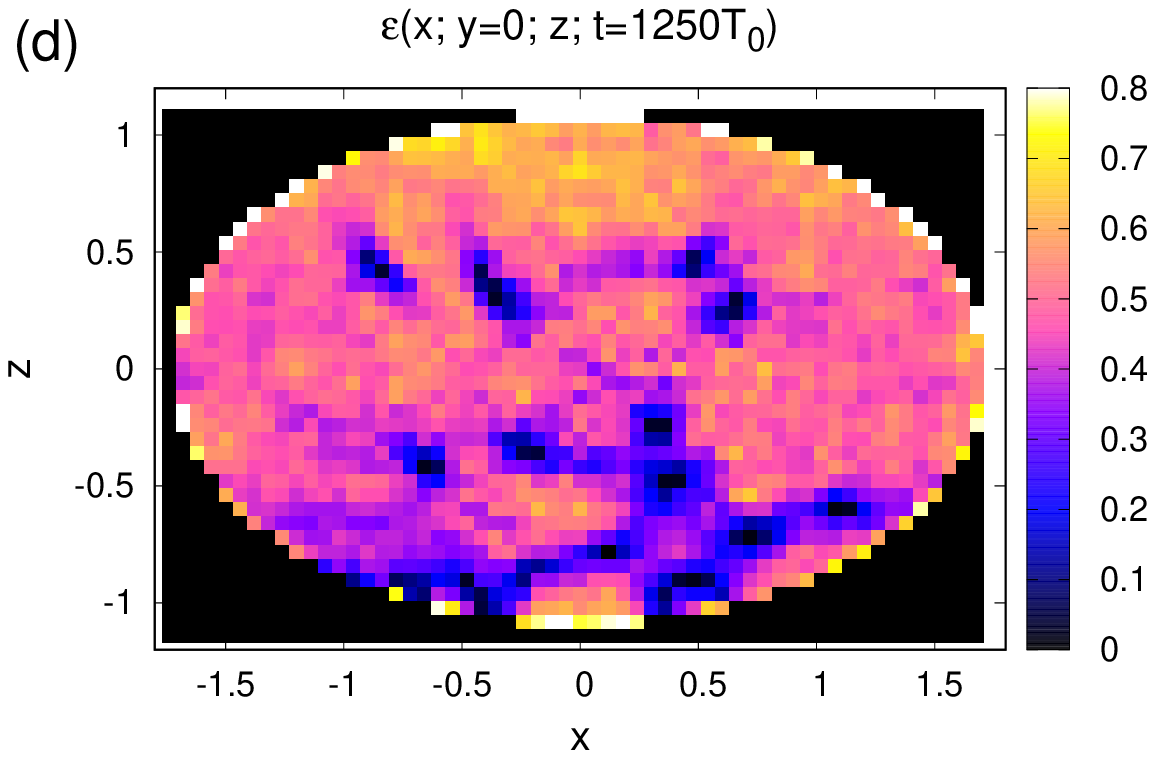, width=75mm}
\end{center}
\caption{
Energies of oscillators in the $y=0$ layer at the parameters 
$\gamma_L=\gamma_C=0.001$, $F=0.12$, and
coupling constant $c$ = (a) 0.02, (b) 0.03, (c) 0.06, and (d) 0.12. 
The background at small $c$ values is not quasi-uniform. 
Vortices in (b) are so thin that their presence is
almost unseen. These figures should be compared with Fig.3a.
}
\label{MU01EQ10C002_12-MU01EQ10C003_12} 
\end{figure}

The dissipative parameters $\gamma_L$ and $\gamma_C$ were for 
simplicity taken the same with an arbitrarily imposed condition 
$\gamma\geqslant 0.001$, which corresponds to the weakened
requirement on the Q factor compared to the freely
damping regime considered in [29].

When using Eq.(3), the detuning of the frequency
of the external signal was taken $\delta=-0.1$. The sign of
detuning is negative in correspondence with the negative 
nonlinear coefficient and negative coupling constants
$c_{n,n'}=-C_{n,n'}/C_0$ (see details in [29, 30]). For
brevity, the symbol $c$ will be used below for the positive
quantity $C_{n,n'}/C_0$.

It is important that, in contrast to systems with
nonlinearity (3), where the $1\rightarrow 2$ decays certainly
occur in the limit of a low energy background at $c>1/4$
and in a significantly nonlinear regime at
smaller $c$ values, three-wave interactions are absent in
the case of symmetric dependence (2). For this reason,
increased parameters are allowed. In particular,
the coupling constant was taken up to $c=2$, detuning
to the extent of $\delta=-0.32$, and damping coefficient up
to $\gamma=0.01$. Vortex filaments were sometimes
observed even at such relatively strong damping.

A slightly perturbed quasi-uniform state was taken
as the initial state. The system ``forgot'' it after several
hundred periods. In this time (at ``favorable'' sets of
the parameters), vortex filaments, often in the form of
rings, nucleated at the boundary of the domain and
then moved to the bulk, where they interacted with
each other. Filaments were deformed, their symmetry
was usually broken, and structures with different
geometry appeared. The picture obviously was not stationary 
because vortices near the axis of the ellipsoid
moved primarily downward, whereas vortices at the
periphery moved upward. However, the characteristic
statistical properties of vortices were mostly quite definite 
and depended strongly on the parameters.

The most typical vortex structures are shown in
Figs. 2-6. Formula (3) and the parameters $h=0.06$
and $\delta=-0.1$ are common for all these figures. Other
parameters are indicated in the figure captions. For a
sufficiently complete visualization of an instantaneous
vortex state, three pictures are usually given. The first
picture shows the energy profile in the $y=0$ section of
the ellipsoid. The second picture presents the quantities 
$\Phi_n=\mbox{arctg}(I_n/V_n)$, which are qualitatively similar 
to the canonical phases $\Theta_n$. The third picture
demonstrates the shape of vortex filaments in the $(x,y)$
projection, where the corresponding $z$ coordinate is
indicated in color.

\begin{figure}
\begin{center}
\epsfig{file=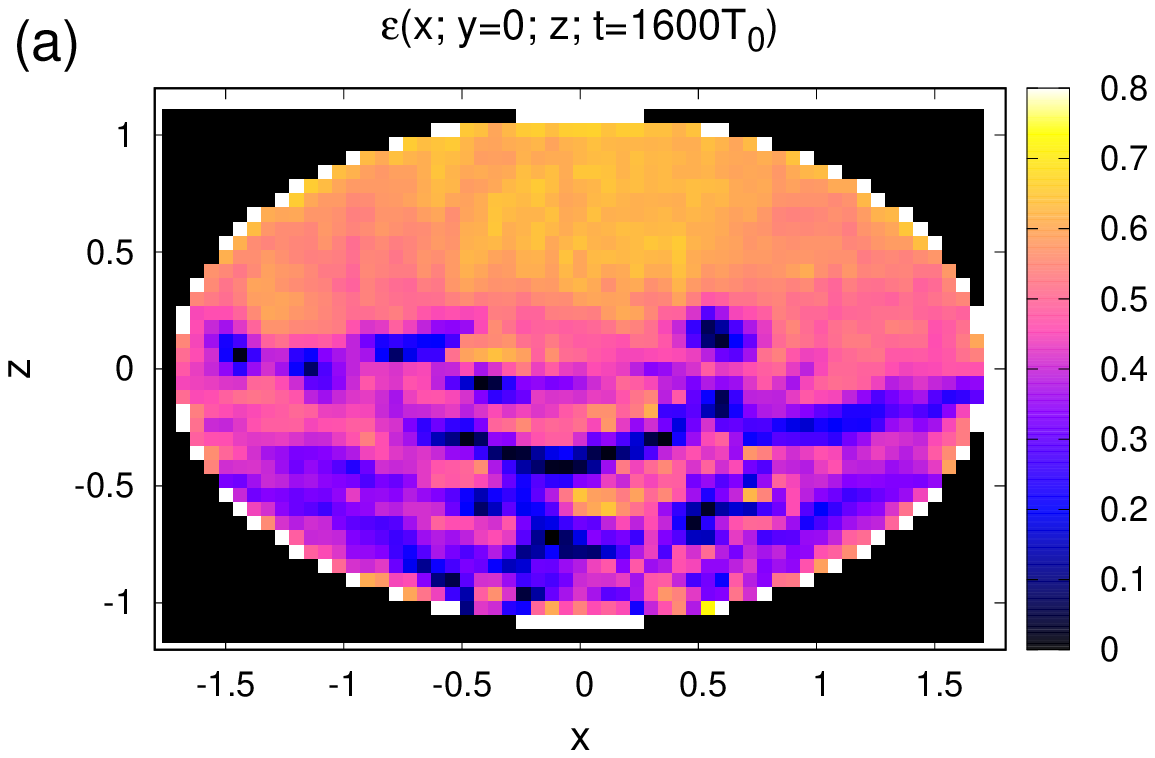, width=75mm}
\epsfig{file=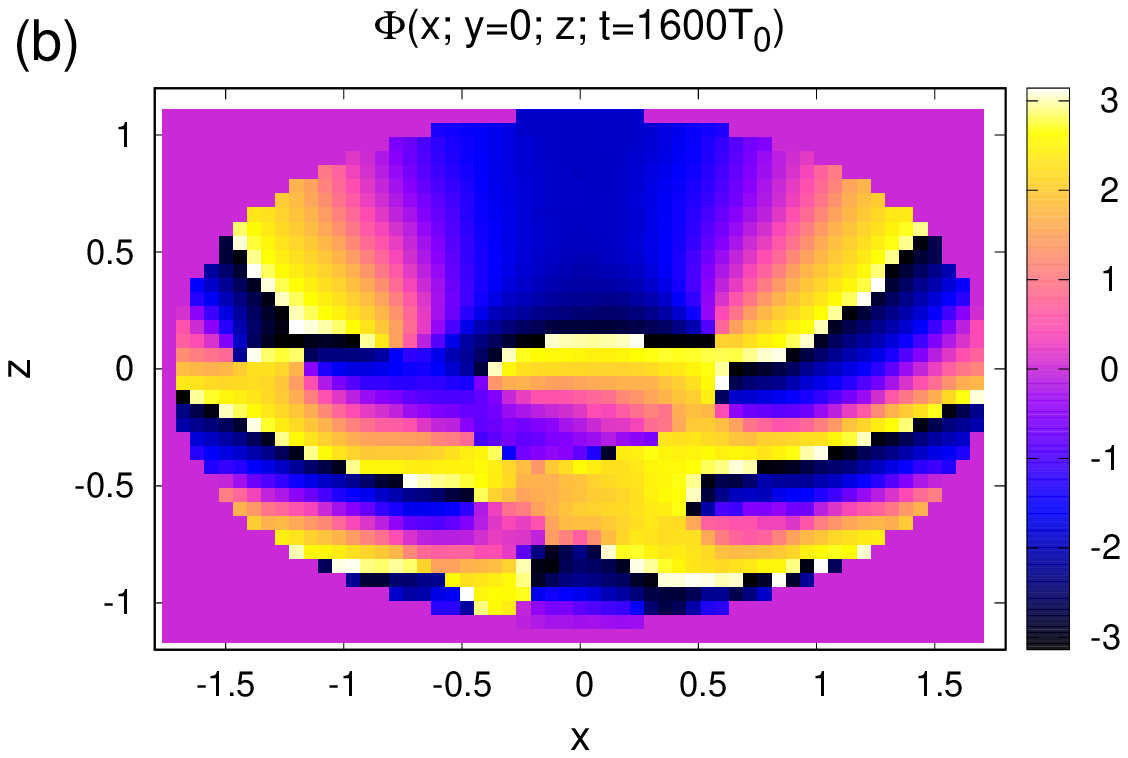, width=75mm}
\epsfig{file=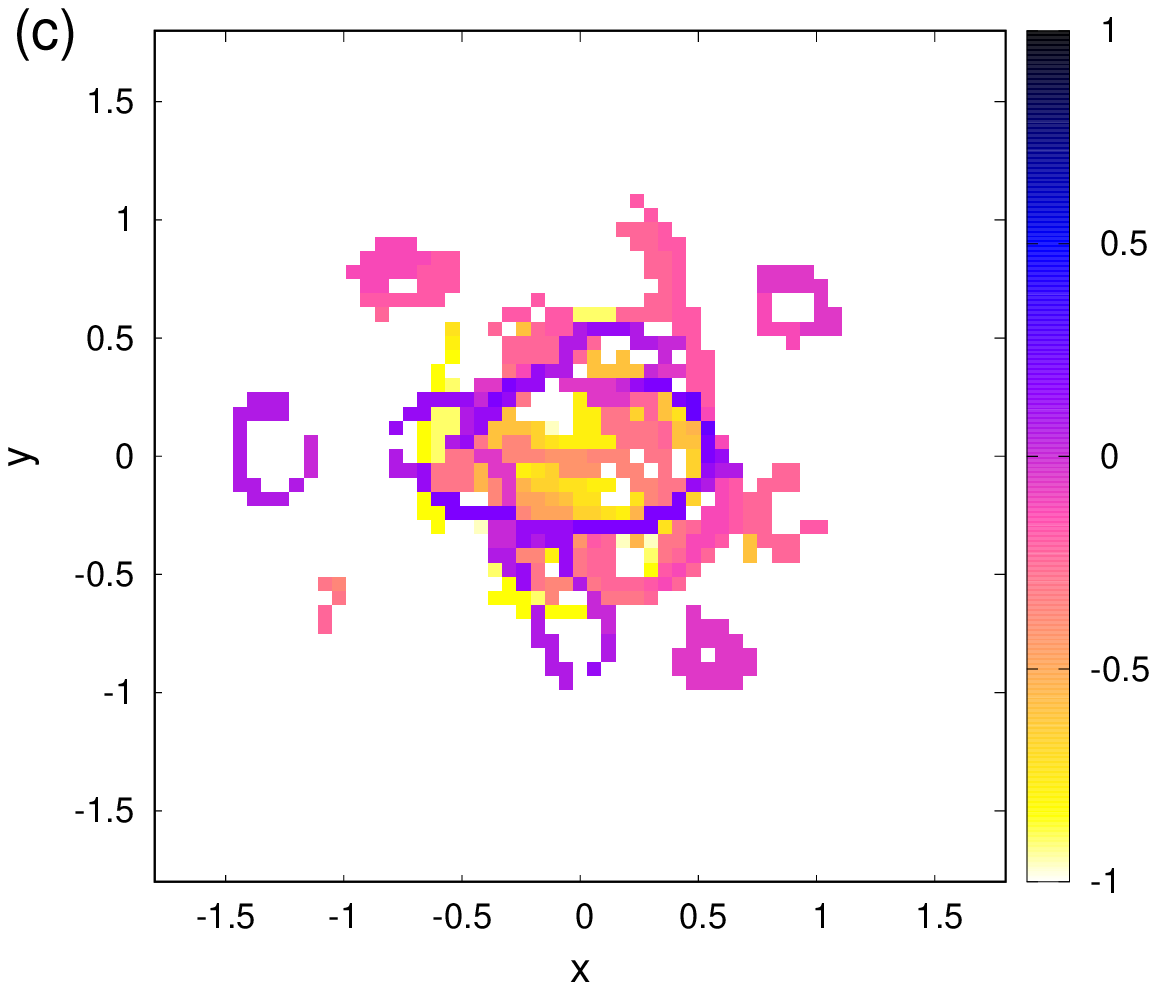, width=75mm}
\end{center}
\caption{
Vortex configuration formed at the parameters 
$\gamma_L=\gamma_C=0.002$, $c=0.1$, and $F=0.20$: (a)
energies of oscillators in the $y=0$ layer, (b) their phases,
and (c) projection of vortex filaments on the $(x,y)$ plane.
Small vortex rings are seen on the periphery.
}
\label{MU02EQ10C01_2} 
\end{figure}
\begin{figure}
\begin{center}
\epsfig{file=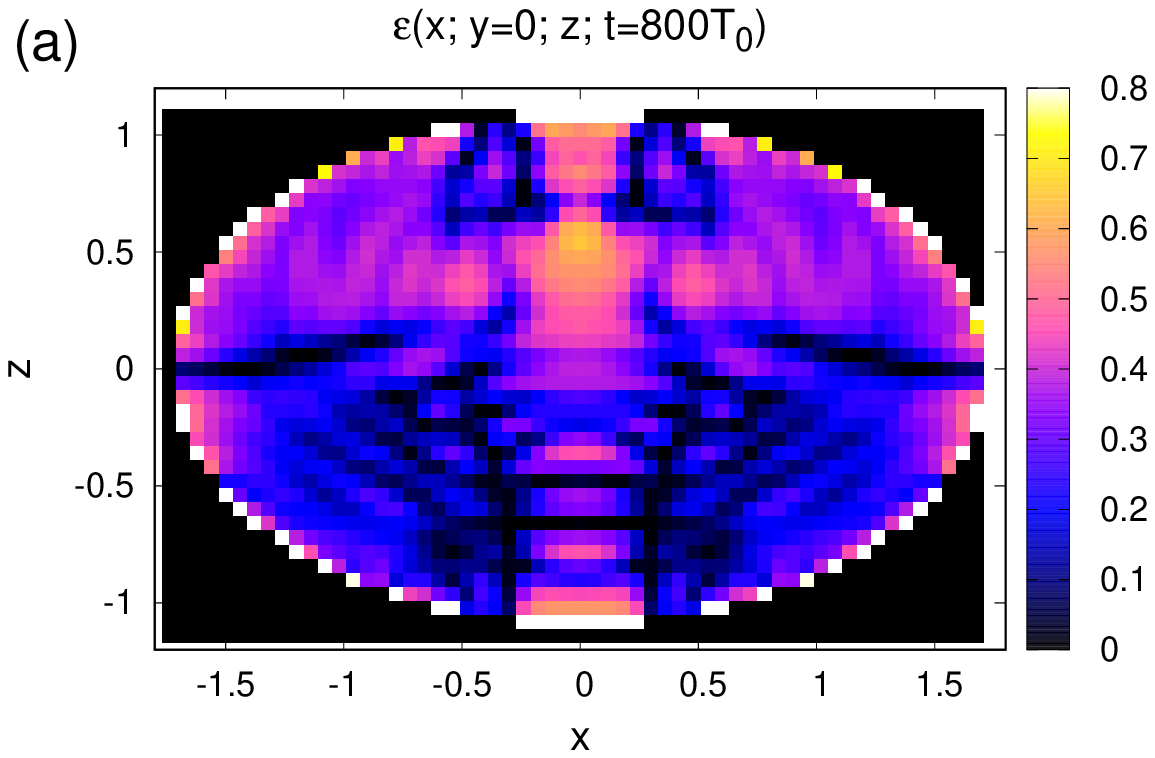, width=75mm}
\epsfig{file=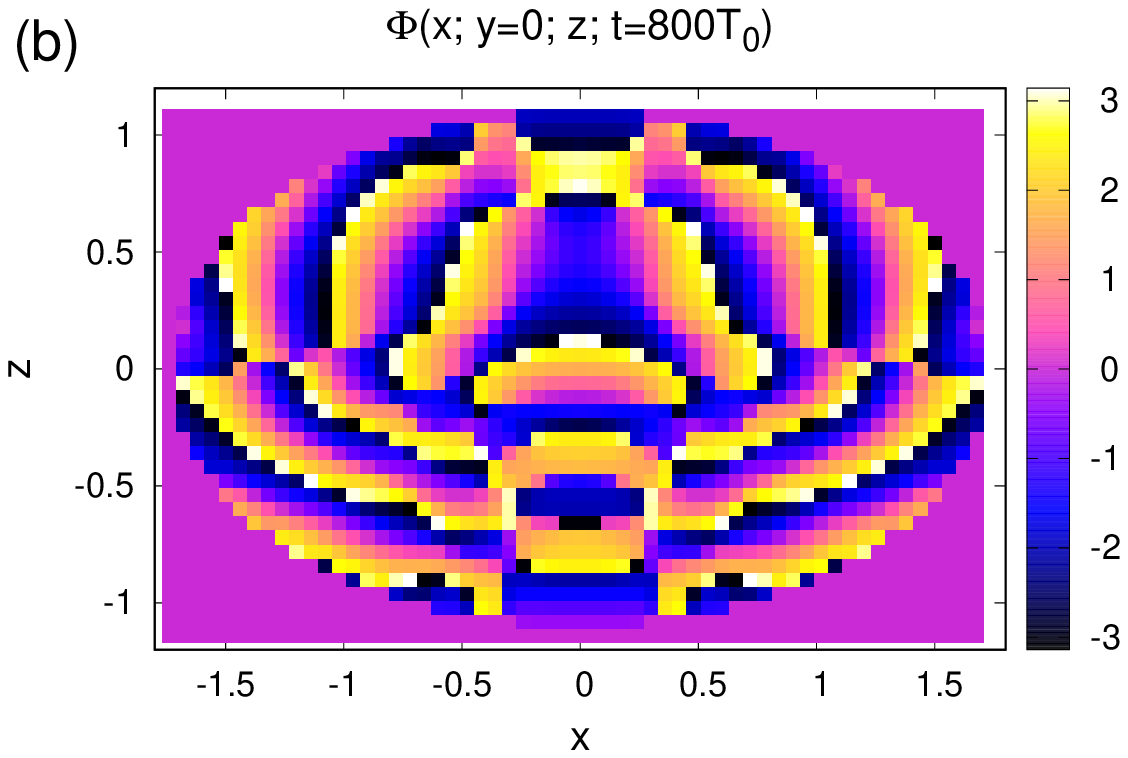, width=75mm}
\end{center}
\caption{
Coherent dissipative structure formed at the parameters
$\gamma_L=\gamma_C=0.006$, $c=0.1$, and $F=0.20$: 
(a) energies of oscillators in the $y=0$ layer and
(b) their phases. The structure consists of vortices in 
combination with dark solitons.
}
\label{MU06EQ10C01_2} 
\end{figure}

A quite developed and incompletely ordered vortex
structure intensely interacting with the boundary is
seen in Fig.2. Such structures appear at moderate
pump amplitudes (in the case under consideration,
$F=0.06$). At smaller amplitudes, the required energy
background is not formed at all (e.g., at $F=0.04$; this
case is not presented in the figures).

When the pump is increased to $F=0.12$, a
``stricter'' and ``quiet'' configuration consisting of
three closely located vortex rings is observed, as shown
in Fig.3. Rings are strongly deformed and are in the
``leapfrogging'' regime. It is noteworthy that similar
structures consisting of four rings were observed at
some other sets of parameters.

Upon increase in $F$ to 0.16 (this case is not illustrated), 
two or three strongly deformed mobile rings at
fairly long distances from each other would be
observed alternatively. At $F=0.20$ (this case is also
not presented in figures), two closely located rings
generally similar to rings at $F=0.12$ would be
observed. The further enhancement of the pump begins 
to damage the background (this case is also not illustrated).

Figure 4 illustrates how the variation of the parameter 
$c$ affects vortex structures. It is seen in Fig.4a that
overly weak links cannot transfer the energy flux sufficient 
to ``fill'' the entire domain taking into account
dissipation. Only a certain pronounced layer near the
boundary is filled with energy, whereas the energy in
the central part is in deficit. It is noteworthy that the
thickness of this layer at a given $c$ value depends primarily
on the dissipative parameter $\gamma$ and slightly on
the pump amplitude. A small increase in the parameter 
$c$ is sufficient to form a quasi-uniform background
(see Fig.4b). In this case, vortices are ``superdiscrete''
because a significant decrease in the energy of oscillators 
near the axis of a vortex hardly occurs. The further
enhancement of links makes cores of vortices thicker
and more noticeable, as seen in Figs.4c and 4d, until
the background begins to be damaged at $c\approx 0.14$ 
(not shown) because of the mentioned $1\rightarrow 2$ 
parametric processes.

Finally, Figs. 5 and 6 demonstrate the evolution of
vortices with an increase in the dissipation rate. In
particular, when $\gamma$ is doubled compared to Figs. 2-4, 
a new feature appears in dynamics: small vortex rings
are produced at the periphery in the bulk of the system
(see Fig.5c), which then drift toward the axis, where
they join the main dissipating structure.

An increase in $\gamma$ to 0.06 fundamentally changes the
entire picture, as shown in Fig.6. In this case, a 
stationary combined structure including dark solitons
appears instead of vortex filaments. The energy background 
is on the whole low and far from uniform. The further 
enhancement of dissipation destroys the background completely.

Thus, a new regime of existence of quantized vortex
filaments in weakly dissipative discrete systems has
been demonstrated. Scenarios of violation of this
regime under the variation of the basic parameters of
the system have been revealed.

\end{document}